\documentstyle[seceq,epsf]{ptptex}
\newcommand{\talk}{\footnote{\sc
       Lectures given by M.I.~Polikarpov                 
   at the 1997 Yukawa International Seminar on           
 "Non-perturbative QCD - Structure of QCD Vacuum -"      
                 (YKIS'97),                              
             2-12 December, 1997,                        
    Yukawa Institute for Theoretical Physics,            
                Kyoto, Japan.                            
\\}}
\newcommand{\itep}{
~\vspace{-.5cm}
\begin{flushright}
{\large ITEP-TH-14/98}
\end{flushright}
\vspace{.5cm}}

\newcommand{\cZ}{{\cal{Z}}}
\newcommand{\cD}{{\cal{D}}}
\newcommand{\cT}{{\cal{T}}}
\newcommand{\cC}{{\cal{C}}}
\newcommand{\tr}{{\rm Tr}}
\newcommand{\dd}{\mbox{d}}
\newcommand{\cO}{{\cal O}}

\newcommand{\dD}{{\cal D}}

\newcommand{\eq}[1]{(\ref{#1})}

\newcommand{\bbbz}{{\mathchoice {\hbox{$\sf\textstyle Z\kern-0.4em Z$}}
{\hbox{$\sf\textstyle Z\kern-0.4em Z$}}
{\hbox{$\sf\scriptstyle Z\kern-0.3em Z$}}
{\hbox{$\sf\scriptscriptstyle Z\kern-0.2em Z$}}}}

\newcommand{\diff}{\partial}
\newcommand{\doublediff}[2]{[\diff_{#1},\diff_{#2}]}

\newcommand{\beq}{\begin{equation}}
\newcommand{\eeq}{\end{equation}}
\newcommand{\beqn}{\begin{eqnarray}}
\newcommand{\eeqn}{\end{eqnarray}}
\newcommand{\bea}[1]{\beq\begin{array}{#1}}
\newcommand{\eea}{\end{array}\eeq}
\newcommand{\summ}[2]{\sum\limits_{#1}^{#2}}

\newcommand{\Z}{{Z \!\!\! Z}}
\title{
\itep
Monopoles in the Abelian Projection of Gluodynamics
\talk
}

\author{Maxim N.~{\sc Chernodub}$^*,$\footnote{E-mail address:
chernodub@vxitep.itep.ru}, Fedor V.~{\sc Gubarev}$^*$,\\
Mikhail I. {\sc Polikarpov}$^*,$\footnote{E-mail address:
polykarp@vxdesy.desy.de}
Alexander I.~{\sc Veselov}$^*$}
\inst{
$^*$ ITEP, B.Cheremushkinskaya 25, Moscow, 117259, Russia
}

\abst{We discuss some properties of the abelian monopoles in compact
$U(1)$ gauge theory and in the $SU(2)$ gluodynamics both on the
lattice and in the continuum.}

\begin{document}

\maketitle

\section{Introduction}

\begin{figure}[htb]
\vskip1cm
\centerline{\epsfxsize=0.5\textwidth\epsfbox{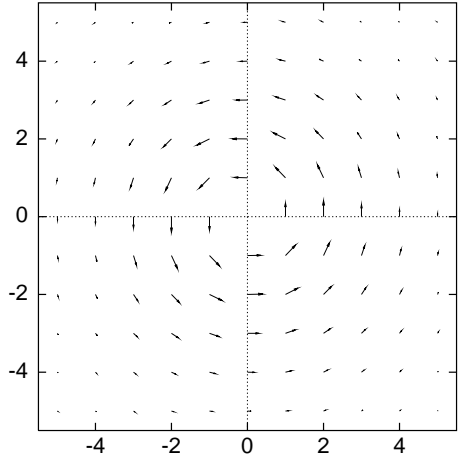}}
\caption{The monopole currents around the string tube which is formed
between the static quark and anti-quark in gluodynamics,
Ref. \cite{bali2}.}
\label{bali2}
\end{figure}

\begin{figure}[htb]
\vspace{-3.5cm}
\centerline{\epsfxsize=.55\textwidth\epsfbox{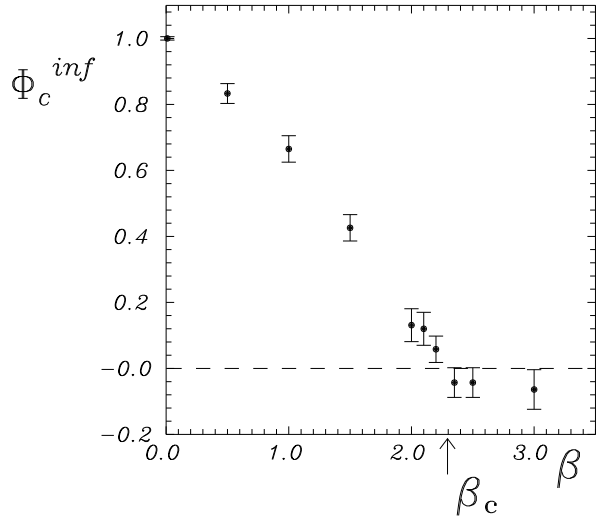}} 
\vspace{0.1cm}
\caption{The dependence of $\Phi_c^{inf}$ on $\beta$, the size of
the lattice is $L^3 \cdot 4$, $L \to \infty$.}
\vspace{-0.6cm}
\label{four}
\end{figure}

Abelian monopoles play a key role in the dual superconductor
mechanism of confinement~\cite{MatH76} in non-abelian gauge theories.
Abelian monopoles appear after the so called abelian
projection~\cite{tHo81}. According to the dual superconductor
mechanism a condensation of abelian monopoles should give rise to the
formation of an electric flux tube between the test quark and
antiquark. Due to a non-zero string tension the quark and the
antiquark are confined by a linear potential. This
mechanism has been confirmed by many numerical simulations of the
lattice gluodynamics \cite{LatRew,ChPo97} which show that the abelian
monopoles in the Maximal Abelian projection are responsible for
the confinement. The $SU(2)$ string tension is well
described by the contribution of the abelian monopole
currents~\cite{MonopoleDominance}; these currents satisfy the London
equation for a superconductor~\cite{SiBrHa93,bali2}. In Fig.~\ref{bali2},
taken from Ref.~\cite{bali2}, the abelian monopole currents near the
center of the flux tube formed by the quark--anti-quark pair are
shown. It is seen that the monopoles wind around the center of the
flux tube just as the Cooper pairs wind around the center of the
Abrikosov string. In Fig.~\ref{four} taken from Ref.~\cite{ChPoVe97}
we show the dependence of the value of the monopole condensate
$\Phi_c^{inf}$ on $\beta$ is shown. It is clearly seen that
$\Phi_c^{inf}$ vanishes at the phase transition and it plays the role
of the order parameters~\cite{ChPoVe97,DiGiAll}. In
Ref.~\cite{SuzEtAl} the effective lagrangian for monopoles was
reconstructed from numerical data for monopole currents for $SU(2)$
gluodynamics in the Maximal Abelian gauge. It occurs that this
lagrangian corresponds to the Abelian Higgs model, the monopole are
condensed in the {\it classical} string tension of the
Abrikosov-Nielsen-Olesen string describes well the {\it quantum}
string tension of the $SU(2)$ gluodynamics. It means that the
description of the gluodynamics at large distances in terms of the
monopole variables can be very useful.

Below we give a short review of the recently obtained results; for
the elementary introduction to the subject see Ref.~\cite{ChPo97}.

Since the abelian monopoles play a dominant role in the confinement
phenomena it is important to understand how do the
abelian monopoles arise in the non-abelian gauge theories. This
question is well understood in the lattice gauge theory while it
deserves some additional study in the continuum theory. In
Section~\ref{origin} we discuss the origin of the abelian monopoles
in the compact electrodynamics.

In Sections~\ref{Polyakov} and~\ref{MaAgauge} we discuss the
procedure of abelian projection in the continuum. As an example we
consider the Polyakov Abelian gauge (Section~\ref{Polyakov}) and the
Maximal Abelian gauge (Section~\ref{MaAgauge}).

In Section~\ref{dyons} we show that the abelian
monopoles carry also an abelian electric charge, this fact means that
the abelian monopoles are dyons. The dyonic nature of the abelian
monopoles is a consequence of the complicated topological structure
of non-abelian vacuum which contains
self-dual non-abelian configurations. Due to the presence of such
configurations the abelian monopoles become abelian dyons.

It is possible that the abelian monopoles correspond to
some non-abelian objects which populate the vacuum of gluodynamics.
If there exists such correspondence one should observe a
non-zero excess of energy density near the monopole world
trajectories.  Indeed, this excess had been found in numerical
simulations of the lattice $SU(2)$ gluodynamics~\cite{BaChPo97}. In
Section~\ref{PhysObjects} we discuss this question in details.

In Section~\ref{3DMaAGauge} we discuss a new three-dimensional
Maximal Abelian projection which is better than the usual Maximal
Abelian projection from the physical point of view.

\section{Siglular Gauge Transformations and Monopoles}
\label{origin}

At first we discuss the question how the abelian monopoles appear in
the $U(1)$ gauge theory. We use the lattice regularization,
the action is a periodic function of the field strength tensor:
\beq\label{Action_U1}
S_{lattice}= \summ{x,\mu>\nu}{} f(Re[U_{x,\mu\nu}])
= \summ{x,\mu>\nu}{} S_{plaq.}(F_{x,\mu\nu})\,;
\qquad U_{x,\mu\nu}=e^{iF_{x,\mu\nu}}\,,
\eeq
where $U_{x,\mu\nu} = U_{x,\mu} U_{x+\mu,\nu} U^{+}_{x+\nu,\mu}
U^{+}_{x,\nu}$ and sum is over all lattice plaquettes.

The obvious requirement is that for the weak fields the lattice
action~(\ref{Action_U1}) is reduced to the continuum action:
\beq
\left.S_{plaq.}\right|_{F_{x,\mu\nu}\to 0}{\rightarrow} F^2_{x,\mu\nu}
\eeq
Since the function $e^{ix}$ is periodic, the
field-strength $F_{x,\mu\nu}$ is physically equivalent to
$F_{x,\mu\nu}+2\pi n_{x,\mu\nu}$,
where $n_{x,\mu\nu}$ is an arbitrary integer-valued lattice two-form.

In the continuum limit $n_{x,\mu\nu}$
is a surface $\delta$-function. Thus in the continuum theory
the gauge fields which correspond to
\beq
F_{\mu\nu}(x) \quad
{\rm and\, to}\quad
F_{\mu\nu}(x)+
2\pi \varepsilon_{\mu\nu\alpha\beta}
\summ{i}{}\Sigma^{(i)}_{\alpha\beta}(x,\tilde{x}^{(i)})
\eeq
are physically equivalent. Here
$\Sigma^{(i)}_{\mu\nu}(x,\tilde{x}^{(i)})$ is an arbitrary
surface $\delta$--function\footnote{This surface is nothing but the
Dirac sheet.}:
\bea{c}
\Sigma^{(i)}_{\mu\nu}(x,\tilde{x}^{(i)})
= \int d^2\sigma^{(i)}_{\mu\nu}\delta^{(4)}(x-\tilde{x}^{(i)}(\sigma^{(i)})),   \\
d^2\sigma^{(i)}_{\mu\nu}=
\varepsilon^{ab}\diff_a \tilde{x}^{(i)}_{\mu}\diff_b
\tilde{x}^{(i)}_{\nu} d^2\sigma^{(i)}
\label{Sdelta}
\eea
the coordinates of the surface, $\tilde{x}^{(i)}_{\mu}
(\sigma^{(i)})$, are parametrized by $\sigma^{(i)}_a, a=1,2$.

Thus defined $U(1)$ gauge theory is invariant under a large class of
the singular gauge transformations, $A_\mu \to A_\mu - i \Omega^+
\partial_\mu \Omega$, where $\Omega=e^{i\alpha}$ is a regular
function (an element of the $U(1)$ group) while the function $\alpha$
may contain discontinuities $\alpha \to \alpha + 2\pi$. Under singular
gauge transformations, $F_{\mu\nu} (A) \to F_{\mu\nu} (A^\Omega)  =
F_{\mu\nu} (A) + [\partial_\mu,\partial_\nu] \alpha$. It may be shown
that $[\partial_\mu,\partial_\nu] \alpha$ corresponds to the surface
$\delta$-function, eq.\eq{Sdelta}. Hence the singular gauge
transformation shifts the field strength tensor as follows:
\beq
F_{\mu\nu}(x) \to F_{\mu\nu}(x)+
2\pi \varepsilon_{\mu\nu\alpha\beta}
\summ{i}{}\Sigma^{(i)}_{\alpha\beta}(x,\tilde{x}^{(i)})\,.
\label{transf}
\eeq
Thus $S[F_{\mu\nu}(A)] = S[F_{\mu\nu}(A^\Omega)]$, where
$S[F_{\mu\nu}(A)]$ is the action of the compact $U(1)$ gauge theory.
The explicit example of such singular gauge transformation is the
following: let $\alpha$ be the azimuthal angle in the polar
coordinate system:
\beq\label{polar}
x_1\pm ix_2=r\sin\gamma e^{\pm i\alpha}\,,
\qquad x_3=r\cos\gamma
\eeq
\bea{c}
F_{\mu\nu}(x)[A^\Omega]=F_{\mu\nu}(x)[A]+
\doublediff{\mu}{\nu}\alpha= \\ \\
=F_{\mu\nu}(x)[A]+
2\pi (\delta_{\mu,1}\delta_{\nu,2}-\delta_{\mu,2}
\delta_{\nu,1}) \delta(x_1)\delta(x_2)
\eea
The singular part in the field-strength appears due to the
topological non-triviality of the gauge matrix, but the action which
corresponds to $\hat{A}_{\mu}$ and to $\hat{A}^\Omega_{\mu}$ is the
same.

Now we discuss how the abelian monopoles appear in the path
integral formalism. In the continuum limit the compact
electrodynamics is described by the following partition function:
\beqn
Z =  \int \dD A \int \dD {\tilde x}
\exp\Bigl\{ - \frac{1}{4 e^2} \int \dd^4 x
{\Bigl(F_{\mu\nu}(A) +  2\pi \varepsilon_{\mu\nu\alpha\beta}
\Sigma_{\alpha\beta}(x,\tilde{x})\Bigr)}^2 \Bigr\}\,,
\label{pf-u1}
\eeqn
this theory is manifestly invariant under the singular gauge
transformations which shift the field strength tensor $F_{\mu\nu}$
according to eq.\eq{transf}. The integration in eq.\eq{pf-u1} is over
collection of all open and closed surfaces $\Sigma$. The simplest
(but not unique) measure of the integration over the surfaces
$\Sigma$ is the string integration measure described in
Ref.~\cite{PolyakovBook}.

It is simple to rewrite the partition function \eq{pf-u1} in the
monopole representation. First we introduce the additional
antisymmetric tensor field $G_{\mu\nu}$ and represent the partition
function as follows:
\beqn
Z & = & \int \dD A \int \dD G \int \dD {\tilde x}
\exp\Bigl\{ - \int \dd^4 x \Bigl(\frac{e^2}{4} G^2_{\mu\nu}
\nonumber\\
& & + i G_{\mu\nu} \bigl( F_{\mu\nu}(A) +  2\pi
\varepsilon_{\mu\nu\alpha\beta} \Sigma_{\alpha\beta}
(x,\tilde{x})\bigr)\Bigr) \Bigr\}\,.
\label{pf-u1-2}
\eeqn
Integrating over the field $A_\mu$ we get a constraint: $\partial_\mu
G_{\mu\nu} = 0$ which can be resolved by the use of the regular
field $B_\mu$: $G_{\mu\nu} = {(2 \pi)}^{-1}
\varepsilon_{\mu\nu\alpha\beta} \partial_\alpha
B_\beta$. Changing the integration in eq.\eq{pf-u1-2} from $G$ to
$B$ we get:
\beqn
Z = \int \dD B \int \dD j
\exp\Bigl\{ - \int \dd^4 x \Bigl(\frac{1}{4 g^2} F^2_{\mu\nu} (B) +
i \, j_\mu(x,{\hat x}) B_\mu \Bigr) \Bigr\}\,,
\label{pf-u1-3}
\eeqn
where $g = 4 \pi \slash e$ is the charge of the monopole, and
the vector ${\hat x}_\mu$ parametrizes the boundaries $j_\mu$
of the Dirac sheets $\Sigma_{\mu\nu}(x,\tilde{x})$:
\beqn
j_\mu = \partial_\nu
\Sigma^{(i)}_{\nu\mu}(x,\tilde{x})\,.
\eeqn
The boundary of the Dirac sheet is the world trajectory of the
monopole. The field $B_\mu$ in eq.\eq{pf-u1-3} plays the role of the
dual gauge field.

Applying the Bardakci-Samuel formula \cite{BaSa78} to eq.\eq{pf-u1-3}
we get:
\beqn
Z = \int \dD B \int \dD \Phi
\exp\Bigl\{ - \int \dd^4 x \Bigl(\frac{1}{4 g^2} F^2_{\mu\nu} (B)
+ \frac{1}{2} {|(\partial_\mu + i B_\mu) \Phi|}^2 \Bigr) \Bigr\}\,,
\label{pf-u1-4}
\eeqn
where $\Phi$ is the (complex) monopole field which carries the
magnetic charge $1$.

The situation with the $SU(2)$ ($SU(N)$) lattice gauge theory is
similar. There are (singular in the continuum) gauge transformations
which do not change the lattice action but create a string--like
singularity in the continuum limit. We discuss the physical
consequences of this fact in the separate publication, below we give
an explicit example of such gauge transformation.

Consider
\beq\label{GaugeMatrix}
\hat{\Omega}(x)=\left[
\begin{array}{cc}
\cos\frac{\gamma}{2} & -\sin\frac{\gamma}{2} e^{-i\alpha}           \\
\sin\frac{\gamma}{2} e^{i\alpha} & \cos\frac{\gamma}{2}
\end{array}\right]
\eeq
where $\alpha$ and $\gamma$ are respectively azimuthal and polar angles
in the polar coordinate system~(\ref{polar}). The action of the
lattice SU(2) gauge model is invariant under this gauge
transformation. But in the continuum,
\beqn
{\hat{F}}_{\mu\nu}({\hat{A}}^{\Omega}) & = &
\Omega^{+} \hat{F}_{\mu\nu}(\hat{A}) \Omega
-i \Omega^{+} \doublediff{\mu}{\nu} \Omega
= \Omega^{+} \hat{F}_{\mu\nu}(\hat{A}) \Omega \nonumber\\
& & - 2\pi (\delta_{\mu,1}\delta_{\nu,2}
- \delta_{\mu,2}\delta_{\nu,1}) \sigma^3
\delta(x_1)\delta(x_2) \Theta(-x_3)
\eeqn
Thus the continuum limit of the lattice action is different from the
naive continuum limit, $\lim\limits_{a\to 0} S_{lattice} \neq \int
d^4x \frac{1}{2}Tr[\hat{F}^2_{\mu\nu}]$, due to the presence of the
singular fields.

\section{Abelian Projection}
\label{Polyakov}

For the sake of simplicity we explain the Abelian projection for
$SU(2)$ gauge group, the generalization to $SU(N)$ gauge group is
straightforward.

The Abelian Projection suggested by 't~Hooft \cite{tHo81} is a
partial gauge fixing defined by the conditions of diagonalization of
some functional $X[A]$ with respect to gauge transformations
$\Omega$, the functional $X[A]$ belongs to the adjoint representation
of the $SU(2)$ gauge group:
\beqn
   X[A] \to X[A^{(\Omega)}] = \Omega^+ X[A] \Omega\,.
\eeqn
In the abelian projection the matrix $X[A]$ is diagonal and the
theory possesses the $U(1)$ gauge symmetry. The generator of this
symmetry is the Cartan group generator $\sigma^3 \slash 2$, where
$\sigma^3$ is the Pauli matrix. The diagonal nonabelian gauge field
$A^3_\mu$ behaves as an abelian gauge field with respect to the
residual $U(1)$ gauge transformations ($A_\mu \to A_\mu - i
\Omega^+_{U(1)} \partial_\mu \Omega_{U(1)}$,
$\Omega_{U(1)} = \exp\{i\sigma^3 \alpha\}$).

The abelian monopoles appear as additional abelian degrees of freedom
which are associated with the singularities in the gauge fixing
conditions. These singularities appear at the points of the space
where eigenvalues of the matrix $X[A]$ coincide.  The proof that
these singularities are monopoles is given in Ref.~\cite{tHo81}. Two
eigenvalues of the matrix $X[A]$ coincide if three independent
equations are satisfied~\cite{tHo81} and in the four-dimensional space
these singularities form closed loops\footnote{The closeness of the
monopole loops reflects the conservation of the magnetic charge. An
explicit proof of this fact is given in Ref.~\cite{KrScWi87}.}.

Consider as an example the Polyakov Abelian projection in $SU(2)$
gluodynamics. This abelian projection is defined for a finite
temperature gauge theory or for the gauge theory in a finite box. The
Polyakov Abelian projection corresponds to the
diagonalization of the functional $P_x$ related to the
Polyakov loop, $P = \frac{1}{2} \tr P_x$:
\beqn
P_x = \cT \exp \left\{ i \oint_{\cC_x} \dd x_0 A_0(x) \right\}\,,
\label{PolLoop}
\eeqn
where the integration is over the closed path $\cC_x$ which
starts and ends in the same point $x$ and is parallel to the "time"
direction; the symbol $\cT$ means the path ordering. The path is
closed due to the periodic boundary conditions.

Now we show that in the continuum all the abelian monopoles are
static in the Polyakov abelian gauge\footnote{The same conclusion was
independently obtained by F.Lenz, private communication.}.  Consider
a monopole trajectory which passes through some point $x$, thus the
eigenvalues of the Polyakov loop coincide with each other in the
point $x$. This means that the matrix $L_x$ belongs to the center of
$SU(2)$ group, $\Z_2$:  $\tr L_x = \pm 2$.  Consider another
point $y$ which lies on the same Polyakov loop (this means that $y_i
= x_i$, $i=1,2,3$). $\tr L_y = \tr L_x = \pm 2$ and the eigenvalues
of the matrix $L_y$ coincide since this matrix belongs to the center
of $SU(2)$ group.  Thus, if the abelian monopole passes through
the point $x=(t_0,\vec{x})$ it also passes through all points $y$
with the same spatial coordinates:  $y=(t,\vec{x})$ for all $t$. Thus
in the Polyakov Abelian projection all abelian monopole trajectories
are static.

Now we discuss the Faddeev--Popov gauge fixing procedure for the
't~Hooft abelian projection. The conditions of diagonalization can be
explicitly written as follows:
\beqn
    C^a[A] = 0\,, \quad
    C^a[A] = \tr \left( X[A] \sigma^a \right)\,, \quad a=1,2\,,
    \label{GaugeConditions}
\eeqn
where $\sigma^a$ are the Pauli matrices. The Faddeev--Popov
determinant $\Delta_{FP}[A]$ is defined by the path integral over the
gauge orbits of $SU(2)$ group:
\beq
  1 = \Delta_{FP}[A] \int \cD \Omega \, \prod_{a=1,2} \delta \left(
  C^a[A^{(\Omega)}] \right)\,.
  \label{FPD-def}
\eeq
Straightforward evaluation of this integral yields:
\beqn
  \Delta_{FP}[A] = const. \, \exp \left\{ 2 \int \dd^4 x
  \ln |\lambda_1[A(x)] - \lambda_2[A(x)]| \right\} \nonumber\\
  = const. \, \exp \left\{ 2 \int \dd^4 x
  \ln |Im\{\lambda_1[A(x)]\}| \right\}\,,
  \label{FPD-lambda}
\eeqn
where $\lambda_a[A(x)]$, $a=1,2$ are the eigenvalues of the matrix
$X[A]$. The Faddeev--Popov determinant \eq{FPD-lambda} is explicitly
gauge invariant.

Substituting unity \eq{FPD-def} into the path integral
\beqn
   \cZ = \int \cD A \, \exp\left\{ - S[A] \right\} \,.
   \label{PF}
\eeqn
and integrating over $\Omega$ we get the partition function in the
abelian gauge \eq{GaugeConditions}:
\beq
   \cZ_{g.f.} = \int \cD A \, \exp\left\{ - S[A] \right\}
   \Delta_{FP}[A]\, \prod_{a=1,2} \delta \left(
   C^a[A] \right)\,.  \label{pf-gf}
\eeq

\section{Maximal Abelian Projection}
\label{MaAgauge}

The most interesting results on abelian monopoles were
obtained in the Maximal Abelian (MaA) gauge. This gauge is defined by
the maximization of the functional
\beqn
  \max_\Omega R[\hat{A}^\Omega]\,,\qquad R[\hat{A}]
  = - \int\, d^4 x \, [(A_\mu^1)^2 + (A_\mu^2)^2]\,,
  \label{R}
\eeqn

The  condition of  a local extremum   of the functional $R$ is
\beqn
(\diff_\mu \pm i g A^3_\mu) A^\pm_\mu =0.
\eeqn
This condition (as well as the functional $R[A]$)
is invariant under the $U(1)$ gauge transformations,
$A_\mu \to A_\mu + \partial_\mu \alpha$. The
meaning of the MaA gauge is simple: by gauge transformations we  make
the gauge field ${\hat A}_\mu$ as diagonal as possible.

The Maximal Abelian gauge on the lattice is defined by the condition
\cite{KrScWi87}:
\beqn
\max_\Omega R[\hat{U}^\Omega_l]\,,\qquad
R[U_l] = \sum\limits_l Tr[\sigma_3 U_l^+ \sigma_3 U_l]\,,
\quad l = \{x,\mu\}\,.
\eeqn
This gauge condition corresponds to an abelian gauge,  since $R$ is
invariant under the $U(1)$ gauge transformations.

Now we discuss the Faddeev--Popov gauge fixing procedure for the MaA
projection in the continuum. We define the Faddeev--Popov unity:
\beqn
1=\Delta_{FP}[A;\lambda] \cdot \int \dD \Omega \, \exp \{ \lambda
R[A^\Omega]\}\,, \quad \lambda \to + \infty\,,
\label{FPU}
\eeqn
where $\Delta_{FP}$ is the Faddeev--Popov determinant. We substitute
the unity \eq{FPU} in the partition function \eq{PF},
shift the fields by the regular transformation $\Omega^+$:
$A \to A^{\Omega^+}$ and use the gauge invariance of the Haar measure,
the action and the Faddeev-Popov
determinant are invariant under the regular gauge transformations.
Thus we get the product of the volume of the gauge orbit,
$\int \dD \Omega$, and the partition function in the fixed gauge:
\beqn
\cZ_{MaA} = \int \dD A \, \exp\Bigl\{- \frac{1}{4}  \int
\dd^4 x \,  F^2_{\mu\nu}[A] + \lambda R[A] \Bigr\}\,
\Delta_{FP}[A;\lambda]\,.
\label{PFMaA}
\eeqn
In the non--degenerate case the FP determinant can be represented in
the form:
\beqn
\Delta_{FP}[A;\lambda] = Det^\frac{1}{2} M[A^{\Omega^{MaA}_r}] \,
\exp\Bigl\{ - \lambda R[A^{\Omega^{MaA}_r}]\Bigr\} + \dots\,,
\eeqn
where $\Omega^{MaA}_r = \Omega^{MaA}_r(A)$ is the
regular gauge transformation which corresponds to a global maximum of the
functional $R[A^\Omega]$,
the dots correspond to the terms which are suppressed in the limit
$\lambda \to \infty$; and
\beqn
M^{ab}_{xy}[A] = \frac{\diff^2 R(A^{\Omega(\omega)})
}{\diff \omega^a (x) \, \diff \omega^b (y)}
{\lower0.15cm\hbox{${\Biggr |}_{\omega=0}$}}\,,
\eeqn
$\Omega(\omega) = \exp\{i\omega^a T^a\}$, $T^a=\sigma^a \slash 2$
are the generators of the gauge group, $\sigma^a$ are the Pauli matrices.
In the limit $\lambda \to + \infty$ the region of the
integration over the fields $A$
reduces to region where the gauge fixing functional $R$ is maximal,
and therefore the partition function \eq{PFMaA} can be rewritten
as follows~\cite{ChPoVe95}:
\beq
        \cZ_{MaA} = \int \dD A \exp\{- S(A)\} Det^{\frac{1}{2}} \Bigl(
        M[A] \Bigr) \Gamma_{FMR}[A]\,,
                \label{GPF2}
\eeq
where $\Gamma_{FMR}[A]$ is a characteristic function of the
Fundamental Modular Region \cite{FMR} for the MaA
projection\cite{ChPoVe95}: $\Gamma_{FMR}[A] = 1$ if the field $A$
belongs to the Fundamental Modular Region (the global maximum of the
functional $R[A]$)  and $\Gamma_{FMR}[A] = 0$ otherwise.

Usually in the abelian projection the $U(1)$ gauge invariant quantities
($\cO$) are
considered. Below we derive the explicit expression for the
 $SU(2)$ invariant quantity $\tilde \cO$ which corresponds to $\cO$.
The expectation value for the quantity $\cO$ in the MaA gauge
\eq{R} is:
\beqn
{<\cO>}_{MaA} =
\frac{1}{\cZ_{MaA}} \int \dD A \, \exp\{ - S(A) + \lambda R[A]\} \,
\Delta_{FP}[A;\lambda] \, \cO(A)\,.
\label{E1}
\eeqn
Shifting the fields $U \to U^{\Omega}$ and integrating over $\Omega$
both in the nominator and in the denominator of expression \eq{E1} we get:
\beqn
{<\cO>}_{MaA} = <{\tilde \cO}>\,,\quad
{\tilde \cO}(A) = \frac{\int \dD \Omega \,
\exp\{ \lambda R[A^\Omega] \} \, \cO(A^\Omega)}{\int \dD \Omega \,
\exp\{ \lambda R[A^\Omega] \}}\,,
\label{tO0}
\eeqn
${\tilde \cO}$ is the $SU(2)$ invariant operator.  In the limit
$\lambda\to + \infty$ we can use the saddle point method to calculate
${\tilde \cO}$:
\beqn
{\tilde \cO}(A) = \frac{\sum\limits^{N(A)}_{j=1}
Det^\frac{1}{2} M[A^{\Omega^{(j)}}] \,
\cO(A^{\Omega^{(j)}})}{\sum\limits^{N(A)}_{k=1}
Det^\frac{1}{2} M[A^{\Omega^{(k)}}]}\,,
\label{tO}
\eeqn
where $\Omega^{(j)}$ are the $N$--degenerate
global maxima of the functional $R[A^\Omega]$
with respect to the regular gauge transformations $\Omega$:
$R[A^{\Omega^{(j)}}] = R[A^{\Omega^{(k)}}]$, $j,k=1,\dots,N$.
In the case of non--degenerate global maximum ($N=1$),
we get ${\tilde \cO}(A) = \cO(A^{\Omega^{(1)}})$.

\section{Abelian Monopoles Carry Electric Charge}
\label{dyons}

Consider a (anti-) self--dual configuration of the $SU(2)$ gauge field:
\beqn
F_{\mu\nu}(A)=\pm \frac{1}{2} \varepsilon_{\mu\nu\alpha\beta}
F_{\alpha\beta}(A) \equiv \pm \tilde{F}_{\mu\nu}\,,
\label{duality}
\eeqn
where $F_{\mu\nu}(A) = \partial_{[\mu,} A_{\nu]} + i [A_\mu,A_\nu]$.
In the MaA projection the commutator term \mbox{${\rm Tr} (\sigma^3
[A_{\mu},A_{\nu}])$} of the field strength tensor $F^3_{\mu\nu}$ is
suppressed, since the MaA projection is defined~\cite{KrScWi87} by
the maximization condition \eq{R}. Therefore, in the MaA projection
eq.(\ref{duality}) yields~\cite{BornSchierholz}:
$f_{\mu\nu}(A)=\partial_\mu A^3_\nu - \partial_\nu A^3_\nu \approx\pm
{\tilde f}_{\mu\nu}(A)$.  Thus, the abelian monopole currents must be
accompanied by the electric currents:  $J^e_{\mu}=\diff_{\nu}
f_{\mu\nu}(A)\approx \pm\diff_{\nu}{\tilde f}_{\mu\nu}(A)=\pm
J^m_{\mu}$. Thus, in the MaA projection the abelian monopoles
are dyons for (anti) self-dual $SU(2)$ field
configurations~\cite{BornSchierholz}. Below we show that in the real
(not cooled) vacuum of lattice gluodynamics the abelian monopole
currents are correlated with the electric currents~\cite{jejm}.

In order to study the relation of electric and magnetic currents, we
have to calculate  connected  correlators of these currents.  The
simplest correlator $\ll J^m_{\mu}J^e_{\mu} \gg \equiv $
$<J^m_{\mu}J^e_{\mu}> - < J^m_{\mu}> <J^e_{\mu}>$ is zero, since
$<J^m_{\mu}J^e_{\mu}> = 0$ due to the opposite parities of the
operators $J^m$ and $J^e$,  and   $<J^{m,e}_{\mu}>=0$ due to the
Lorentz invariance. The simplest non--trivial (normalized) correlator
is
\beqn
{\bar G } =
\frac{1}{\rho^e \rho^m} \, <J^m_{\mu}(y) J^e_{\mu}(y) q(y)>\,,
\label{G}
\eeqn
where $q(x)$ is the sign of the topological charge density at the
point $x$ and $\rho_{m,e} = $ $ \sum_{l} <|J^{m,e}_l|> \slash (4V)$
are the densities of the magnetic and the electric charges, $V$ is
the lattice volume (total number of sites).

\begin{figure}[ht]
\begin{center}
\centerline{{\epsfxsize=0.5\textwidth\epsfbox{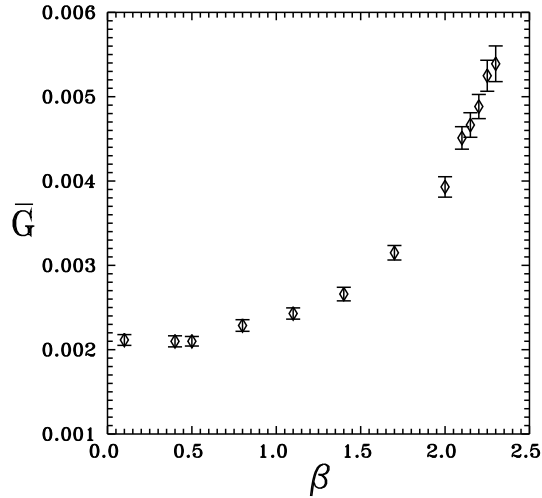}}}
\end{center}
\vskip-5mm
\caption{The dependence of the correlator ${\bar G}$ on $\beta$}
\label{one}
\end{figure}

We perform the numerical calculation of the  correlator \eq{G} in the
$SU(2)$ lattice gauge theory on the $8^4$ lattice with periodic
boundary conditions. We use  100 statistically independent gauge
field configurations for each value of $\beta$.

The dependence of the correlator ${\bar G}$ on $\beta$ is shown in
Fig.~\ref{one}. This correlator is positive for  all values of
$\beta$. Therefore, the abelian monopoles in the MaA projection carry
an electric charge. According to definition \eq{G}, the sign of the
electric charge of the monopole coincides with the product of the
magnetic charge and the topological charge. Thus, in the gluodynamic
vacuum the abelian monopoles become abelian dyons due to a
non-trivial topological structure of the vacuum gauge
fields~\cite{jejm}.

\section{Abelian Monopole Currents are Correlated with $SU(2)$
Action Density}
\label{PhysObjects}

Abelian monopoles appear as singularities in the gauge
transformations~\cite{tHo81,LatRew,ChPo97}
(see also Sections 2-4). On the
other hand, the monopole currents reproduce the $SU(2)$ string
tension~\cite{MonopoleDominance}. Thus, monopoles  are likely to be
related to some physical objects. A physical object is something
which carries action. Below we study the local correlations of the
abelian monopoles with the density of the magnetic and the electric
parts of the $SU(2)$ action (the global correlation was found in
Ref.~\cite{ShSu95}). We show that the monopoles are physical objects
but it does not mean that they have to propagate in the Minkowsky
space; a chain of instantons can produce a  similar effect: an
enhancement of the action density along a line in  Euclidean space.
The simplest quantities which can show this correlation are the
relative excess of the magnetic and the electric action densities
$\eta^{M,E} = (S^{M,E}_m - S) \slash S$ in the region near the
monopole current.  Here $S$ is the expectation value of the lattice
plaquette action, $S_P = <( 1 - \frac 12 Tr \, U_{P})>$. The
quantities $S^{M,E}_m$ are, respectively, the magnetic and the
electric parts of the $SU(2)$ action density, which are calculated on
plaquettes closest to the monopole current.

\begin{figure}[h]
\begin{center}
\centerline{{\epsfxsize=0.5\textwidth\epsfbox{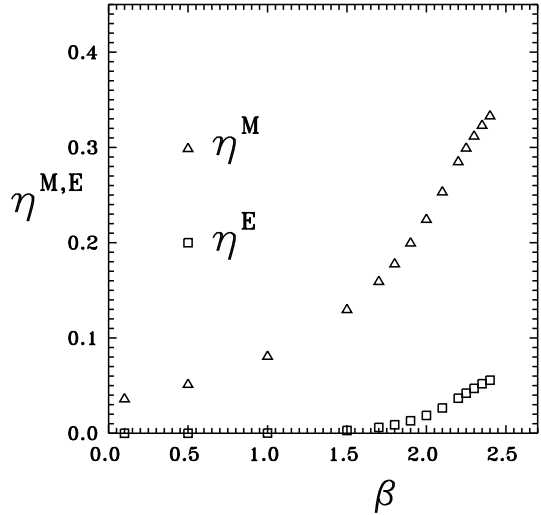}}}
\end{center}
\vskip-5mm
\caption{The relative excess of the magnetic (triangles, from
Refs.~\cite{BaChPo97}) and the electric (boxes) action density near
the monopole current.}
\label{two}
\end{figure}

In the continuum notation, the quantities $S^{M,E}_m$ have the following
form:
\beqn
S^M_m   =   \frac{1}{2} <{\rm Tr}{(
n_\mu (x)\,{\tilde F}_{\mu\nu}(x))}^2>\,,\quad
S^E_m   =   \frac{1}{2} <{{\rm Tr} {(n_\mu(x)\,
F_{\mu\nu}(x)})}^2 >\,,
\eeqn
$n_\mu (x)$ is the unit vector in the direction of the current:
$n_\mu (x)=j_\mu (x)/| j_\mu (x) |$, if $j_\mu (x) \neq 0$, and
$n_\mu (x)= 0$ if $j_\mu (x) = 0$. It is easy to see that for a
static monopole ($j_0 \neq 0; \,\, j_i=0, i=1,2,3 $) $S^M_m$
($S^E_m$) corresponds to the chromomagnetic action density
${(B^a_i)}^2$ (chromoelectric action density ${(E^a_i)}^2$) near
the monopole current.

We calculate the quantities $\eta^{M}$ and $\eta^{E}$ on the lattice
$24^4$ with periodic boundary conditions. In Fig.~\ref{two} we show
the quantities $\eta^{M,E}$ $vs.$ $\beta$. The monopole currents are
calculated in the MaA projection. In Fig.~\ref{two} the statistical
errors are smaller than the size of the symbols. It is clearly seen
that the abelian monopoles are correlated with the magnetic and
the electric parts of the $SU(2)$ action density. Note that the
correlation of the monopole current with the magnetic action density
is larger than the correlation of the monopole current with the
electric action density.

\begin{figure}[t]
\vskip5mm
\centerline{{\epsfxsize=0.5\textwidth\epsfbox{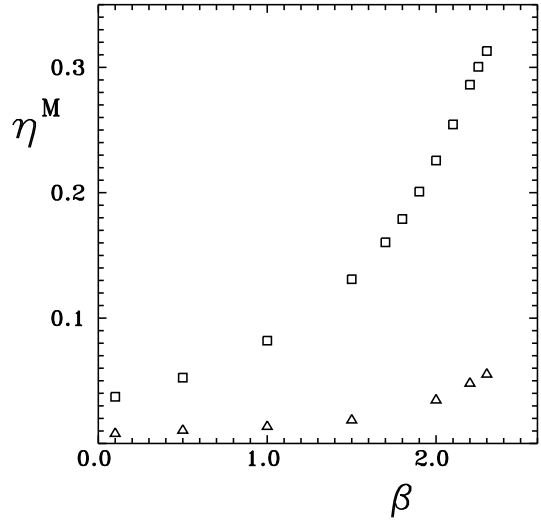}}}
\caption{The relative excess of the magnetic
action density near the monopole current for $1^3$
(squares) and $2^3$ (triangles) monopoles on $24^4$ lattice.}
\label{three}
\end{figure}

The similar results are obtained with extended
monopoles~\cite{IvPoPo90} which are defined on the cubes of the
size $N \times N \times N$~\cite{BaChPoVe98}. In
Fig.~\ref{three} we show the quantities $\eta^M$ $vs.$ $\beta$
for the extended monopoles of the sizes $n^3$, $n=1,2$ on the lattice
$24^4$.

\section{3D Maximal Abelian Gauge and Effective Monopole Potential}
\label{3DMaAGauge}

As we already discussed the largest part of the numerical calculations is
performed in the MaA projection. Usually the expectation values
of abelian operators (operators constructed from the abelian gauge fields,
diagonal gluons) are calculated in this projection. But
abelian operators correspond to nonlocal in time operators in terms of
the original $SU(N)$ fields $U_{x,\mu}$ (see eq. \eq{tO}).
This nonlocality occurs since the gauge fixing condition \eq{R}
contains time like links $U_{x,4}$. For time nonlocal operators there
are obvious problems with the transition from the Euclidean to
Minkowsky space--time.  Thus there are problems with physical
interpretation of the results obtained for abelian operators in the
MaA projection.

\begin{figure}[t]
\vskip5mm
\begin{tabular}{cc}
\hspace{-3mm}
{\epsfxsize=0.43\textwidth\epsfbox{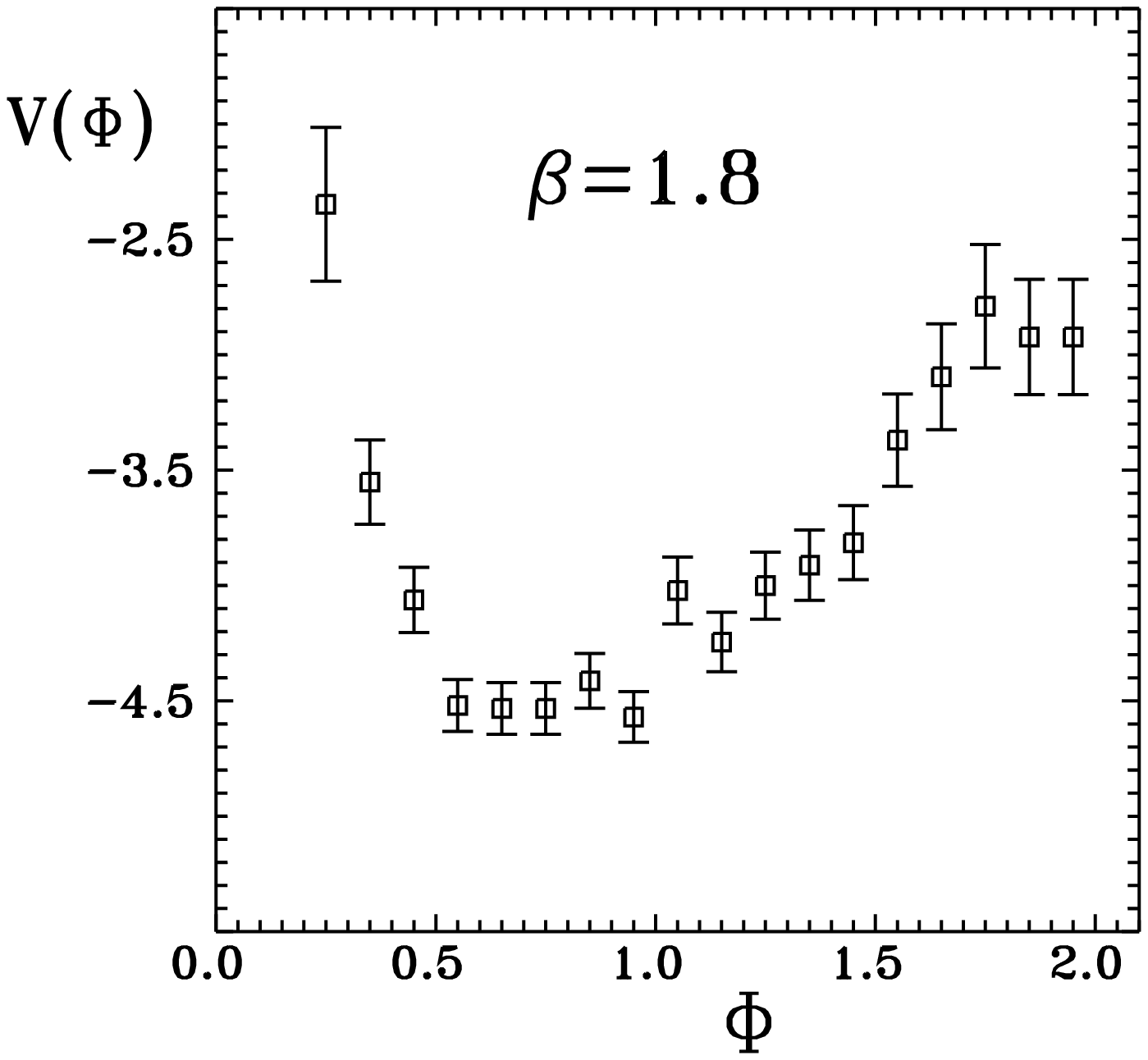}}
&
\hspace{3mm} {\epsfxsize=0.43\textwidth\epsfbox{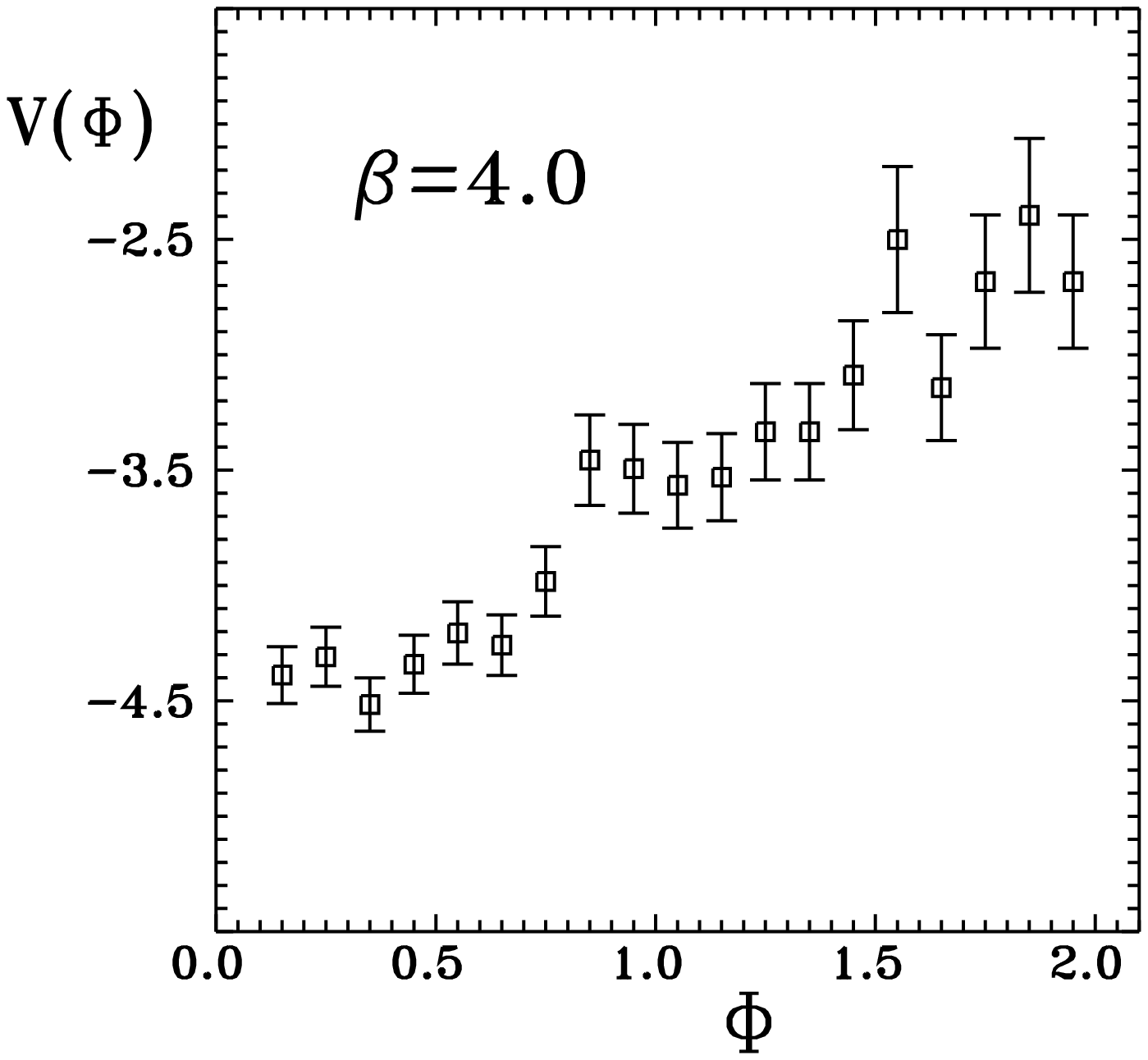}}
\\
~~~(a) & ~~~~~~(b)\\
\end{tabular}
\caption{The effective potential for the monopole fields $\Phi$
(for $\Phi > 0$) in the confinement~(a) and the deconfinement~(b)
phases for the $3D$ MaA gauge.}
\label{onefig}
\end{figure}

Below we present first results of calculations in the 3D MaA projection
which is defined by the same maximization condition as the usual MaA
projection \eq{R}, but  the summation in $R$ \eq{R} is over the
space--like links\footnote{This gauge was discussed by U.-J.~Wiese in
1990, was recently rediscovered by D.~Zwanzinger (private
communication to M.I.P.), and discussed by M.~M\"uller-Preussker at
this school.}. Since the time--like links are excluded from the gauge
fixing condition, the abelian operators in the 3D MaA projection
correspond to local in time operators constructed from the nonabelian
fields. In ref. \cite{ChPoVe97} the effective potential for the
monopole creation operator was calculated in the MaA projection. As
we discussed in the Introduction in the confinement region (below the
critical temperature) this potential is of the Higgs type, above the
critical temperature this potential has minimum at the zero value of
the monopole field. Also this behavior of the potential is very
reasonable it is important to prove the monopole condensation in
the 3D MaA gauge. In Fig.~\ref{onefig}  (a,b) we show the effective
monopole potential for the confinement phase ($\beta = 1.5$, $12^3
\cdot 4$ lattice) and for the deconfinement phase ($\beta = 2.5$,
$12^3 \cdot 4$ lattice). It is clearly seen that the minimum of the
potential is at the nonzero value of the monopole field for the
confinement phase and is at the zero value for the deconfinement
phase. Our definition of the effective potential $V(\Phi)$ is the
same as in Ref.~\cite{ChPoVe97}:  \beq e^{-V(\Phi)} = <\delta (\Phi -
\Phi_{mon}(x))>, \eeq here $\Phi_{mon}(x)$ is the monopole creation
operator, defined in ref.\cite{ChPoVe97}. We discuss the dependence
of the minimum of the effective potential on the temperature for
the 3D MaA projection in a separate publication.

\section*{Acknowledgments}

The authors are grateful to F.~Lenz, M.~M\"uller-Preussker, A. van
der Sijs, Yu.A.~Simonov, T.~Suzuki, V.I.~Zaharov and D.~Zwanziger for
useful discussions.

This work was supported by the grants INTAS-96-370,
INTAS-RFBR-95-0681, RFBR-96-02-17230a  and RFBR-96-15-96740.


\begin{thebibliography}{99}

\bibitem{MatH76} S.~Mandelstam,
\JL{Phys.~Rep.,23C,1976,245}.\\
G.~{'t~Hooft}, "High Energy Physics", {\rm A.~Zichichi, Editrice
Compositori, Bolognia}, 1976.

\bibitem{tHo81} G.~'t~Hooft,
\JL{Nucl.~Phys.,B190 [FS3], 1981,455}.

\bibitem{LatRew} T.~Suzuki, {Nucl.~Phys.} {\bf B} {
(Proc.~Suppl.)} {\bf 30} (1993) 176.\\
M.~I.~Polikarpov, {Nucl.~Phys.}
{\bf B} {(Proc.~Suppl.)} {\bf 53} (1997)
134.

\bibitem{ChPo97} M.N.~Chernodub and M.I.~Polikarpov,
Lectures given at NATO Advanced Study Institute on Confinement,
Duality and Nonperturbative Aspects of QCD, Cambridge, England, 23
June - 4 July 1997, {\tt hep-th/9710205}.

\bibitem{MonopoleDominance} H.~Shiba and T.~Suzuki,
\JL{Phys.~Lett.,B333,1994,461}.\\
J.D.~Stack, S.D.~Neiman and R.J.~Wensley,
\JL{Phys.~Rev.,D50,1994,3399}.\\
G.~Bali, V.~G.~Bornyakov, M.~M\"uller-Preussker and
K.~Schilling, \JL{Phys.~Rev.,D54,1996,2863}.

\bibitem{SiBrHa93}
V.~Singh, D.~Browne and R.~Haymaker, \JL{Phys.~Lett.,B306,1993,115}.

\bibitem{bali2} Ch.~Schlichter, G.S.~Bali and K.~Schilling,
{\tt hep-lat/9709114}.\\
Lectures given by K. Schilling at this School, {\tt hep-lat/9802005}.

\bibitem{ChPoVe97} M.N.~Chernodub, M.I.~Polikarpov and A.I.~Veselov,
\JL{Phys.~Lett.,B399,1997,267}.

\bibitem{DiGiAll} L.~Del Debbio, A.~Di~Giacomo, G.~Paffuti and
P.~Pieri, \JL{Phys.~Lett.,B355,1995,255}.\\
N.~Nakamura,
V.~Bornyakov, S.~Ejiri, S.~Kitahara, Y.~Matsubara and T.~Suzuki,
\JL{Nucl.~Phys.~Proc.~Suppl.,53,1997,512}.\\
Lectures given by A.Di~Giacomo at this School, {\tt hep-lat/9802008}.

\bibitem{SuzEtAl}
S.~Kato, M.N.~Chernodub, S.~Kitahara, N.~Nakamura, M.I.~Polikarpov,
T.~Suzuki, {\it preprint KANAZAWA-97-15}, {\tt hep-lat/9709092}.\\
M.N. Chernodub, S.~Kato, S.~Kitahara, N.~Nakamura, M.I.~Polikarpov and
T.~Suzuki, in preparation.\\
Lectures given by T.~Suzuki at this School.

\bibitem{BaChPo97} B.L.G.~Bakker, M.N.~Chernodub and M.I.~Polikarpov,
\JL{Phys.~Rev.~Lett.,80,1998,30}.\\
B.L.G.~Bakker, M.N.~Chernodub and M.I.~Polikarpov, {\tt hep-lat/9709038}.

\bibitem{PolyakovBook} A.M.~Polyakov, {\it ``Gauge Fields and Strings''}
(Harwood, New York, 1987).

\bibitem{BaSa78}
K.~Bardakci and S.~Samuel, \JL{Phys.~Rev.,D18,1978,2849}.

\bibitem{KrScWi87} A.S.~Kronfeld, M.L.~Laursen, G.~Schierholz and
U.J.~Wiese, \JL{Phys.~Lett.,198B,1987,516}.\\
A.S.~Kronfeld, G.~Schierholz and U.J. Wiese,
\JL{Nucl.~Phys.,B293,1987,461.}

\bibitem{ChPoVe95} M.N.~Chernodub, M.I.~Polikarpov and A.I.~Veselov,
\JL{Phys.~Lett.,B342,1995,303}.

\bibitem{FMR}
V.N.~Gribov, \JL{Nucl.~Phys.,B139,1978,1}.\\
M.A.~Semenov--Tyan--Shanskii and  V.A.~Franke, Proc. Seminars of the
Leningrad Math. Instit. (1982), english translation (Plenum, New
York, 1986).\\
D.~Zwanziger, \JL{Nucl.~Phys.,B378,1992,525}.\\
D.~Zwanziger, \JL{Nucl.~Phys.,B399,1993,477}.\\
D.~Zwanziger, Lectures given at this School, {\tt hep-th/9802180}.

\bibitem{BornSchierholz}
V.Bornyakov, G.Schierholz, \JL{Phys.~Lett.,B384,1996,190}.

\bibitem{jejm} M.N. Chernodub, F.V. Gubarev and M.I. Polikarpov,
{\tt hep-lat/9709039}.\\
M.N. Chernodub, F.V. Gubarev and M.I. Polikarpov, {\tt hep-lat/9801010},
to published in Phys.~Lett. {\bf B}.

\bibitem{ShSu95} H.~Shiba and T.~Suzuki, \JL{Phys.~Lett.,B351,1995,
519}

\bibitem{IvPoPo90} T.L.~Ivanenko, A.V.~Pochinskii and M.I.~Polikarpov
\JL{Phys.~Lett.,B252,1990,631}.

\bibitem{BaChPoVe98} B.L.G.~Bakker, M.N.~Chernodub, M.I.~Polikarpov
and A.I.~Veselov, to be published.

\end{thebibliography}
\end{document}